\documentclass[preprint,showpacs,preprintnumbers,amsmath,amssymb]{revtex4}

\usepackage{graphicx}
\usepackage{dcolumn}
\usepackage{bm}
\usepackage{epsfig}

\newcommand{\z}{&&\hspace*{-1cm}}

\newcommand{\bea}{\begin{eqnarray}}
\newcommand{\eea}{\end{eqnarray}}
\newcommand{\be}{\begin{equation}}
\newcommand{\ee}{\end{equation}}

\begin{document}

\boldmath
\title{Gluon density
from the Berger-Block-Tan form of the  
structure function $F_2$}
\unboldmath

\author{N.Yu.~Chernikova}
\affiliation{Sunday school,
141980 Dubna (Moscow region), Russia.
}
\author{
A.V. Kotikov
\footnote{E-mail address: kotikov@theor.jinr.ru}
}
\affiliation{
Bogolubov Laboratory for Theoretical Physics, JINR,
141980 Dubna (Moscow region), Russia.}

\date{\today}

\begin{abstract}
We present a set of formulas to extract the gluon density
from the Berger-Block-Tan form of the deep inelastic structure function 
$F_2$
at small $x$ in the leading
order of perturbation theory.

\end{abstract}

\pacs{11.10.Hi, 11.15.Me, 12.38.-t, 12.38.Bx}
\keywords{Parton distribution functions, deep-inelastic structure functions}
\maketitle

For experimental studies of hadron-hadron processes on the 
LHC collider, it is necessary to know in detail the
values of the parton (quark and gluon) distribution functions
(PDFs) of nucleons, especially at small values of Bjorken variable $x$. The basic
information about PDF properties can be 
extracted from the process of deep inelastic
lepton-hadron scattering (DIS). 

In the snall-$x$ regime, nonperturbative effects were expected to give important
contributions. However,
what is observed up to very low $Q^2 \sim 1$ GeV$^2$ values, traditionally explained by soft processes,
is described reasonably well by perturbative OCD (pQCD) evolution (see, for example, \cite{CooperSarkar:1997jk}).
The pQCD evolution leads to a ruther singular PDF behavior at small $x$ values
(see \cite{Kotikov:1998qt} and references therein), which violates the Froissard boundary \cite{Froissart:1961ux}.

Recently the new form of the DIS structure function (SF) $F_2(x,Q^2)$ was proposed in \cite{Berger:2007vf}, which 
will be called below as the Berger-Block-Tan (BBT) structure function. The SF  $F_2^{\rm BBT}(x,Q^2)$ 
leads to the low $x$ 
asymptotics of the (reducted) DIS cross-sections $\sim \ln^2 1/x$, which is in turn in an agreement with
the Froissard predictions
\cite{Froissart:1961ux}.

  In the present paper we study the behaviour of the gluon density $xf_g(x,Q^2)$
at small values of $x$, using 
the SF
$F_2^{\rm BBT}(x,Q^2)$ and following to our previous studies in \cite{Kotikov:1993xe}.
We will show
that, the small $x$ behaviour of the corresponding gluon density $xf_g^{\rm BBT}(x,Q^2)$
can be extracted directly from the 
values of $F_2^{\rm BBT}(x,Q^2)$ (see Eq. (\ref{n9}) below) at the leading order (LO) of perturbation theory. 

{\bf 1.}
At the LO the SF  $F_2$
relates with the quark density $sf_q(x,Q^2)$ ss
\be
F_2(x,Q^2) = e xf_q(x,Q^2),~~~ e=\frac{1}{f} \sum^f_{i=1} e_i^2 \equiv \frac{e_{2f}}{f} \, , 
\label{F2}
\ee
where $e$ is the average charge square
and $f$ is the number of active quarks.

The famous DGLAP equations \cite{Gribov:1972ri}
relate the gluon and quark densities at LO as
 \be
\frac{d (xf_a(x,Q^2))}{dlnQ^2}  = -\frac{a_s(Q^2)}{2} \sum_{a,b=s,g} P^{(0)}_{ab}
(x) \otimes  xf_b(x,Q^2),
\label{DGLAP} 
\ee
where $a_s(Q^2)$ the strong coupling constant
\be
a_s(Q^2) = \frac{\alpha_s(Q^2)}{4\pi} = \frac{1}{\beta_0\ln(Q^2/\Lambda^2_{\rm LO})}
\label{as}
\ee
and 
$P^{(0)}_{ab} (x)$ $(a,b=s,g)$ and  $\beta_0$ are the LO splitting functions and the first coefficient of QCD $\beta$-function.
The symbol $\otimes$ marks the 
 Mellin convolution
\bea
f_1(x) \otimes f_2(x) \equiv \int^1_x \frac{dy}{y} 
 f_1(y) f_2\left(\frac{x}{y}\right) \, .
\label{Mellin}
\end{eqnarray}

Using the above equations (\ref{F2}) and (\ref{DGLAP})
together we have got the following final seт of equations
\bea
\z    \frac{dF_2(x,Q^2)}{dlnQ^2}  = -\frac{a_s(Q^2)}{2} \biggl[
e 
P^{(0)}_{sg} (x) \otimes xf_g(x,Q^2) + 
P^{(0)}_{ss} (x)  \otimes  
F_2(x,Q^2)  
 \biggr] ,
\label{dF2} \\
\z 
\frac{d (xf_g(x,Q^2))}{dlnQ^2}  = -\frac{a_s(Q^2)}{2}  \biggl[
P^{(0)}_{gg} (x) \otimes xf_g(x,Q^2) + e^{-1} 
P^{(0)}_{gs} (x)  \otimes   
F_2(x,Q^2)  
 \biggr] 
\label{fg} \, .
\eea

Following to the BBT form $F_2^{\rm BBT}(x,Q^2)$ in (\ref{n9}), 
the first equation can be considered as the definition of the gluon density $xf_g^{\rm BBT}(x,Q^2)$.

The second equation is usually violated beyond
the perturbation theory. The most often
violation is done by addition of the term $\sim (xf_g)^2$ into the r.h.s. of (\ref{fg}) (see, foe example,
Refs.\cite{Gribov:1984tu,Zhu:1998hg}). This
term modifies strongly the low $x$ asymptotics of the gluon density, which becomes to be ruther flat.
Such types of gluon density leads to less singular form of the high-energy asymptotics of cross-sections
\cite{Fiore:2004nt}.
Below,
we will not take into account
the equation (\ref{fg}), but we think that it is violated in such way to obtain
the BBT-like
asymptotics $\sim \ln^2(1/x)$ for $xf_g^{\rm BBT}(x,Q^2)$ and $F_2^{\rm BBT}(x,Q^2)$ at small $x$ values.

{\bf 2.}
The Mellin convolution, shown above in (\ref{Mellin}), tells us that 
the Mellin transform of the
equation (\ref{dF2}) simplifies strongly its form. Indeed, after the Mellin transform
the r.h.s. will contain only simple products of the
corresponding Mellin moments. So, we go to Mellin space, using the following Mellin moments
\bea
&&M_2(n,Q^2) = \int^1_0 dx x^{n-2} F_2(x,Q^2),~~~
M_g(n,Q^2) = \int^1_0 dx x^{n-1} f_g(x,Q^2)\, ,
\label{Mg}\\
&&\gamma^{(0)}_{ab} (n) = \int^1_0 dx x^{n-2} 
P^{(0)}_{ab} (x) \, .
\label{gamma}
\end{eqnarray}

After the Mellin transform, the
equation (\ref{dF2}) leads to the following form
\be
\frac{dM_2(n,Q^2)}{dlnQ^2}  = -\frac{a_s(Q^2)}{2} \biggl[
e \gamma^{(0)}_{sg} (n) M_g(n,Q^2) + \gamma^{(0)}_{ss} (n)  
M_2(x,Q^2)  
 \biggr] ,
\label{dM2} 
\ee
with the  LO anomalous dimensions $\gamma^{(0)}_{sp} (n)$ 
($p=s,g$), which have the following form
 \be
\gamma^{(0)}_{sg} (n) = - \frac{4f (n^2+n+2)}{n(n+1)(n+2)},~~~
\gamma^{(0)}_{ss} (n) ~=~ \frac{32}{3} \biggl[\Psi(n)-\Psi(1)
-\frac{3}{4} + \frac{1}{2n} + \frac{1}{2(n+1)}\biggr],  
\label{ana}
\ee
where $\Psi(n)$ is the Euler $\Psi$-function.

{\bf 3.}~~ We use the extended  BBT form of the
proton SF
at $x < x_P=0.011$, following to Ref. \cite{Block:2011xb}, where the combine H1 and ZEUS data 
\cite{Aaron:2009aa} taken into account:
\be
F_{2}^{\rm BBT}(x,Q^2) =
(1-x) \, \sum_{m=0}^2 A_m(Q^2) L^m  
\label{n9}
\ee
where
\bea
A_0 &=& \frac{F_P}{1-x_P},~~~ F_P=0.413 \pm 0.003,~~~ 
L= \ln \left[\frac{(1-x) x_P}{(1-x_P)x}\right]
\nonumber \\
A_i(Q^2) &=& \sum_{k=0}^2 a_{ik} \, l^k = \sum_{k=0}^2 \tilde{a}_{ik} \, \ln^k(Q^2),~~
i=(1,2),~~ l= \ln (Q^2/\Lambda^2_{\rm LO})
\label{n9.0}
\eea
with
\be
a_{i2}=\tilde{a}_{i2},~~a_{i1}=\tilde{a}_{i1}-2q \tilde{a}_{i2},
~~a_{i0}=\tilde{a}_{i0}-q \tilde{a}_{i1}+q^2 \tilde{a}_{i2},~~
q=-2\ln(\Lambda_{\rm LO}),
\label{n9.1}
\ee
and
\bea
&&\tilde{a}_{10}  \cdot 10^{2} = -8.471 \pm 0.252,~~
\tilde{a}_{11} \cdot 10^{2} = 4.180 \pm 0.156,~~
\tilde{a}_{12} \cdot 10^{4} = -3.973 \pm 2.139, \nonumber \\
&&\tilde{a}_{20} \cdot 10^{2} = 1.297 \pm 0.030,~~
\tilde{a}_{21} \cdot 10^{3}=
2.473 \pm 0.246,~~
\tilde{a}_{22} \cdot 10^{3} =
1.642 \pm 0.055 \, .
\label{n10}
\eea

We will try to find the gluon density $f_g^{\rm BBT}(x,Q^2)$ in a similar form
\footnote{Notice that we omit the factor $(1-x)$, which contributes into Eq. (\ref{n9}).
It is in agreement with the quarks sum rules, where $f_s(x)/f_g(x) \sim (1-x)$ at $x \to 1$.}
\be
f_g^{\rm BBT}(x,Q^2) = \sum_{m=0}^2 B_m(Q^2) L^m,~~~
B_i(Q^2) = \sum_{k=0}^2 b_{ik} \,  l^k \, , 
i=(0,1,2),~~
\label{n9.1}
\ee

Performing the Mellin transforms (\ref{Mg}) and (\ref{gamma}), we have the following representations
for the
$M_k(n,Q^2)$ $(k=2,g)$ at $n=1+\omega$ and  $\omega \to 0$
\bea
M_{2}^{\rm BBT}(n,Q^2) &=& \frac{x_p^{\omega}}{\omega}
\biggl[ A_0 + \frac{1}{\omega} A_1(Q^2) 
+ \frac{2}{\omega^2}  A_2(Q^2) 
+ O(x_p)
\biggr] 
\label{M2BBT}\\
\z \nonumber \\
M_{g}^{\rm BBT}(n,Q^2) &=& \frac{x_p^{\omega}}{\omega}
\biggl[ B_0+ \frac{1}{\omega} B_1(Q^2) 
+ \frac{2}{\omega^2} B_2(Q^2) 
+ O(x_p)
\biggr] \, .
\label{MgBBT}
\eea

These results are based on the 
basic integrals
\be \int^{x_p}_0 dx x^{n-2} (1-x) L^k(x) = k!\frac{x_p^{\omega}}{\omega^{k+1}} +  O(x_p^{\omega+1})
\label{Int}
\ee

From Eqs.(\ref{dM2}),
(\ref{M2BBT}) and (\ref{MgBBT}) we have
 \bea    \z   
A'_1(Q^2) \, L +
A'_2(Q^2) \, L^2  ~=~ -\frac{a_s}{2} \biggl[ \nonumber \\
\z e \biggl\{
\gamma^{(0)}_{sg}  B_0(Q^2) + 
\left(\gamma^{(0)}_{sg} L + \dot{\gamma}^{(0)}_{sg}
\right) B_1(Q^2) + \left(\gamma^{(0)}_{sg} L^2 + 2 \dot{\gamma}^{(0)}_{sg} L +
\ddot{\gamma}^{(0)}_{sg} \right) B_2(Q^2)  \biggr\} \nonumber \\
\z + 
\biggl\{
\gamma^{(0)}_{ss}  A_0 + \left(\gamma^{(0)}_{ss} L + \dot{\gamma}^{(0)}_{ss}
\right) A_1(Q^2) + \left(\gamma^{(0)}_{ss} L^2 + 2 \dot{\gamma}^{(0)}_{ss} L +
\ddot{\gamma}^{(0)}_{ss} \right) A_2(Q^2)  \biggr\} 
 \biggr],
\label{9.4} 
\eea
where 
\bea
\z A'_i(Q^2) =  \frac{A_i(Q^2)}{dlnQ^2} = 
 a_{i}^{(1)} +  2 a_{i}^{(2)} \, l, \label{9.5} \\
\z \gamma^{(0)}_{sp} = \gamma^{(0)}_{sp}(1),~~
\dot{\gamma}^{(0)}_{sp} = 
\frac{d}{d\delta}  \gamma^{(0)}_{sp}(1+\delta)|_{\delta=0},~~
\ddot{\gamma}^{(0)}_{sp}  = 
\frac{d^2}{(d\delta)^2}  \gamma^{(0)}_{sp}(1+\delta)|_{\delta=0}.
\label{9.6} 
\eea

Comparing the l.h.s. and r.h.s. at same power of $L$ we see that
\bea
\z B_2(Q^2) = - \frac{2}{e a_s \gamma^{(0)}_{sg}} A'_2(Q^2), ~~
B_1(Q^2) = - \frac{2}{e a_s \gamma^{(0)}_{sg}} A'_1(Q^2)
- \frac{2\dot{\gamma}^{(0)}_{ss} }{e  \gamma^{(0)}_{sg}} A_2(Q^2)
-\frac{2\dot{\gamma}^{(0)}_{sg}}{ \gamma^{(0)}_{sg}} B_2(Q^2), \nonumber \\
\z B_0(Q^2) = - \frac{1}{e \gamma^{(0)}_{sg}} 
\biggl(\dot{\gamma}^{(0)}_{ss} A_1(Q^2) + \ddot{\gamma}^{(0)}_{ss} A_2(Q^2)
\biggr)
- \frac{1}{\gamma^{(0)}_{sg}} 
\biggl(\dot{\gamma}^{(0)}_{sg} B_1(Q^2) + \ddot{\gamma}^{(0)}_{sg} B_2(Q^2)
\biggr),
\label{9.7} 
\eea
where we used the propoerty $ \gamma^{(0)}_{ss}(1)=0$.

Using the above values (\ref{ana}) for the anomalous dimensions, we have
\bea
\gamma^{(0)}_{sg} &=& - \frac{8f}{3},~~~
\dot{\gamma}^{(0)}_{sg} ~=~ \frac{26f}{9},~~~
 \ddot{\gamma}^{(0)}_{sg} ~=~ - \frac{172f}{27}, \nonumber \\
\gamma^{(0)}_{ss} &=& 0, ~~~
  \dot{\gamma}^{(0)}_{ss} ~=~ \frac{32}{3} \, I_1,~~~
 \ddot{\gamma}^{(0)}_{ss} ~=~ \frac{32}{3} \, I_1 ,  
\label{9.8} 
\eea
with $I_1=\zeta(2)-5/8$ and $I_2=9/8-2\zeta(3)$. Putting the results (\ref{9.8}) to Eqs. (\ref{9.7}), we obtain
\bea
\z B_2(Q^2) =  \frac{3}{4f} \frac{1}{ea_s} A'_2(Q^2),~~
B_1(Q^2) =  \frac{3}{4f} \frac{1}{ea_s} A'_1(Q^2)
+  \frac{8}{ef} \, I_1 A_2(Q^2)
+\frac{13}{6} B_2(Q^2), \nonumber \\
\z B_0(Q^2) =  \frac{4}{ef} \biggl( I_1 A_1(Q^2) +  I_2 A_2(Q^2) \biggr) 
+\frac{13}{12} B_1(Q^2) - \frac{89}{36} B_2(Q^2),
\label{9.9} 
\eea

Comparing the lhs and rhs parts, we find
the coefficients $b_{ik}$ shown in (\ref{n9.1})
\bea
\z b_{20} = 0,~~ 
 b_{2j} = \frac{3j\beta_0 }{4 e_{2f}} a_{2j},~~
b_{10} = \frac{8I_1}{ e_{2f}} a_{20},~~ 
b_{1j} = \frac{3j\beta_0}{4 e_{2f}}a_{1j} + \frac{8 I_{1j}}{ e_{2f}}\, a_{2j}, \nonumber \\
\z 
b_{00} = \frac{4I_{1}}{ e_{2f}}\, a_{10} + \frac{4 \tilde{I}_2}{ e_{2f}}\, a_{20},~~
b_{01} = \frac{4I_{11}}{ e_{2f}}\, a_{11} + \frac{4I_{21}}{ e_{2f}}\, a_{21},~~
b_{02} = \frac{4I_{1}}{ e_{2f}}\, a_{12} +  
\frac{4I_{22}}{ e_{2f}}\, a_{22},
\label{9.11} 
\eea
where $j=(1,2)$ and 
\be
I_{1j}=I_1+ \frac{13j\beta_0}{64},~~\tilde{I}_2 = I_2 + \frac{13}{6} I_1,~~I_{2j}=\tilde{I}_2- \frac{3j\beta_0}{128} \, .
\label{9.12} 
\ee


{\bf Conclusion}
Using the suggested form (\ref{n9.1}) for the gluon density $f_g^{\rm BBT}(x,Q^2)$, we recovered its parameters
at the LO
of perturbation theory
as combination of the corresponding parameters (\ref{n9.0}) of the SF $F_{2}^{\rm BBT}(x,Q^2)$.

The gluon density $f_g^{\rm BBT}(x,Q^2)$ 
is significantly less singular in comparison with the corresponding PDFs in perturbation theory (see, for
example, \cite{Kotikov:1998qt}). Such form is in an agreement with
the Froissard boundary \cite{Froissart:1961ux} and 
can be useful for study \cite{Fiore:2004nt,Arguelles:2015wba}
of the high-energy behavior of photon, neutrino and proton cross sections.

As the next step of our investigations we plan extend the analysis beyond the leading order (see \cite{Kotikov:1994vb}). 
Moreover we plan 
to reconstract the longitudinal SF $F_L$ at low $x$ values following to Ref. \cite{Kotikov:1994jh}.

\begin{acknowledgments}
  A.V.K.
was supported in part by the RFBR Foundation through Grant No.\
16-02-00790-a. 
\end{acknowledgments}

\end{document}